\documentstyle[epsfig]{aipproc}
\begin{document}

\rightline{Report. No. Damtp-2000-128, HD-THEP-00-57, IFIC-00-69}
\vskip -0.1in

\title{Large-Scale Primordial Magnetic Fields from Inflation and Preheating
       \footnote{Talk presented by O. T\"ornkvist at  the conference
{\it Cosmology and Astroparticle Physics (CAPP-2000)} in Verbier, Switzerland 
(July 17-28, 2000).}}

\author{Ola T\"{o}rnkvist$^{*}$\footnote{Short-term visitor,
CERN Theory Division, 1211 Gen\`{e}ve 23, Switzerland},
Anne-Christine Davis$^*$, Konstantinos
Dimopoulos$^\ddag$, and Tomislav Prokopec$^\S$}
\address{$^*$Department of Applied Mathematics and Theoretical Physics,
University of Cambridge\\
Wilberforce Road, Cambridge CB3~0WA, United Kingdom\\
$^\ddag$Instituto de F\'{\i}sica Corpuscular,
Universitat de Valencia/CSIC,\\
Apartado de Correos 2085, 46071 Valencia, Spain\\
$^\S$
Institut f\"ur Theoretische Physik der Universit\"at Heidelberg\\
Philosophenweg 16, D-69120 Heidelberg, Germany}

\maketitle

\vspace*{-7mm}

\begin{abstract}
We consider models in which the (transverse) photon mass
is non-zero during inflation and
drops to zero non-adiabatically at the end of inflation. Through this
process, vacuum fluctuations of the photon field are converted into
physical, long-wavelength modes with high amplitude. The resulting spectrum
of the field strength is approximately $B_\ell\sim \ell^{-1}$, where $\ell$
is the relevant coherence scale. With a reasonable model of field evolution
we obtain, on comoving galactic scales, a magnetic field strong enough to seed
a dynamo mechanism and generate the observed galactic magnetic fields.
\end{abstract}

\vskip -0.1in

\section*{Introduction}

Spiral galaxies typically contain magnetic fields
with strength of about  $10^{-6}$ G  that are aligned with
the spiral density waves \cite{Kronberg}.
A plausible explanation is that these fields have been produced via
a galactic dynamo mechanism, in which a weak seed field
was amplified by the differential
rotation of the galaxy in conjunction with magnetohydrodynamic turbulence.
The required strength of the seed field is subject to large uncertainties;
for a flat, critical-density universe,
estimates lie in the range $10^{-23}$--$10^{-19}$ G at the time
of completed galaxy formation. This lower bound
can be relaxed \cite{dlt} to about $10^{-30}$ G for a
universe with a dark-energy component (e.g. a cosmological constant or
quintessence), which appears to be favoured by recent results from
supernova observations and balloon experiments.

For the dynamo to work, the galactic
seed field should be correlated on a scale of
$\sim 100$ pc, corresponding to the largest turbulent eddy.
Because the magnetic field strength $B$ and correlation length $\ell$
scale as $B\sim \ell^{-2}\sim\rho^{2/3}$ during the
collapse of matter into a galaxy, where $\rho$ is the matter density, the
required comoving correlation length is $\sim 10$ kpc \cite{dlt}.
On this scale, the lower
bounds for the seed field are $2\times 10^{-27}$ G for a critical-density
universe, and $2\times 10^{-34}$ G for a flat, dark-energy dominated,
low-density universe.

Many proposals have been put forward regarding the
origin of the seed field \cite{Grassoandme}. Among the most interesting
is the possibility that the field is primordial and was generated in the
early Universe through one of several field-theoretic mechanisms that
occur in models of particle physics. Such a mechanism must occur out of
equilibrium, which limits the choice to either inflation or a phase
transition. In phase transitions, 
however, it is difficult
to explain the large correlation length.
For instance, the comoving scale corresponding to the {\em horizon\/}
at any time
before the last (QCD) phase transition is much smaller than 10 kpc,
so the seed field is unlikely to have been produced by {\em causal}
 processes in a phase transition in the early universe.

Inflationary models solve problems with causality (such as the horizon
problem) by stretching space itself, without changing the speed of light rays
relative to a local, inertial frame.
For example, it is well known that
long-wavelength vacuum
fluctuations of scalar fields grow during inflation and
provide
the seeds for large-scale structure in the form of density perturbations
\cite{Mukhanov}.
In a similar way, one might expect
inflation to be able to produce magnetic fields correlated on scales that are
beyond causality bounds.

\vspace*{-1mm}
\section*{Evolution of gauge fields in inflation}

In a conformal metric
$ds^2=a^2(\tau) [d\tau^2-d\vec{x}^2]$ (which characterises de Sitter inflation
and flat FRW eras), gauge fields $A_\mu$
are conformally invariant
and do not couple gravitationally. In fact, the equation of motion can be
reduced to $\Box A_\mu=0$, where $\Box=\eta^{\mu\nu}\partial_\mu\partial_\nu$
is the Minkowski-space d'Alembertian. This equation has only
harmonic solutions,
and thus the amplitude of vacuum fluctuations
of $A_\mu$ is a constant function of (conformal) time $\tau$.

In order to have a changing amplitude of vacuum fluctuations, one must somehow
break conformal invariance \cite{tw}. A simple way would be to add, by
hand, a mass term $m^2 g^{\mu\nu} A_\mu A_\nu/2$ to the Lagrangian.
This gives rise to the following equation for a (transverse)
Fourier mode ${\cal A}_{\vec{k}}$\,:
\begin{equation}
\left(\partial^2_\tau + k^2 +\frac{m^2}{H_{\rm I}^2\tau^2}\right)
{\cal A}_{\vec{k}} = 0~,
\label{modeeq}
\end{equation}
where we have used that the scale factor $a=-1/H_{\rm I}\tau$
 in de~Sitter
inflation with Hubble parameter $H_{\rm I}$.
It is obvious that the addition of such a positive mass term will not lead
to growth in the amplitude ${\cal A}_{\vec{k}}~$; on the contrary, the
amplitude
of long-wavelength vacuum fluctuations
decreases during inflation.

To see this in detail, we investigate the exact solutions of
Eq.~(\ref{modeeq}), which are
\begin{equation}
{\cal A}_{\vec{k}}^{(i)} = \frac{1}{2} \sqrt{-\pi\tau}
H_\nu^{(i)}(-k\tau)~,\quad\quad i=1,2,
\quad\quad \nu=\frac{1}{2}
\sqrt{1 - \frac{4 m^2}{H_{\rm I}^2}}~,
\label{sol}
\end{equation}
where
$H^{(i)}_\nu$ are Hankel functions.
Towards the beginning of inflation ($\tau\to -\infty$),
the solutions behave as  $(2k)^{-1/2} \exp(\pm i k\tau)$ [up to a phase]
such that
$|{\cal A}_{\vec{k}}^{(i)}|\sim (2k)^{-1/2}$ and $|\partial_\tau {\cal
A}_{\vec{k}}^{(i)}|
\sim (k/2)^{1/2}$. At the end of inflation, which in our parametrisation occurs
at $\tau=-1/H_{\rm I}$, we have for long wavelengths ($k/H_{\rm I} \ll 1$)
that $|{\cal A}_{\vec{k}}^{(i)}|\sim (2k)^{-1/2} \Gamma(\nu)(k/2H_{\rm
I})^{1/2-\nu}$
and
 $|\partial_\tau {\cal A}_{\vec{k}}^{(i)}|\sim
(k/2)^{1/2} \Gamma(\nu)(1-2\nu)(2H_{\rm I}/k)^{1/2+\nu}/4$.
Because $\nu<1/2$, the amplitude has decreased. Note, however, that the
electric-field amplitude
$|\partial_\tau {\cal A}_{\vec{k}}^{(i)}|$
has increased
enormously for very long wavelengths $k\ll H_{\rm I}$. This is a consequence
of the Heisenberg uncertainty relation. We shall see next how such a large
electric field is converted into a magnetic field when the mass $m$ suddenly
drops to zero at the end of inflation.

\vspace*{-3mm}
\section*{Non-adiabaticity}

In order to understand the concept of non-adiabaticity, consider a
quantum particle in the ground state of a
one-dimensional infinite well. If the walls of the
well are suddenly moved
apart (so quickly that the particle has no time
to exchange energy with its surroundings),
the particle
will find itself in a quantum
state which is, in general, a linear superposition of excited states of the
new, larger well. In the case of a quantum field,
such an excited state corresponds to
a physical observable field, distinct from the vacuum.

In our scenario \cite{ddpt},
we propose that the photon mass $m$ goes non-adiabatically
to zero at the end of inflation, so that the massive-photon mode
functions described
by Eq.~(\ref{sol}) are suddenly projected onto mode functions $(2k)^{-1/2}
\exp(\pm i k \tau)$
that are solutions of the massless equation. In the quantum
language, this can be expressed as a Bogoliubov transformation
$\tilde{a}_{\vec{k}}
= \alpha_{\vec{k}} a_{\vec{k}} + \beta_{\vec{k}}^{\ast}
a_{\vec{k}}^\dagger$
between
annihilation operators $\tilde{a}_{\vec{k}}$ and $a_{\vec{k}}$ for the quantum
fields with $m\neq 0$ and $m=0$, respectively. The values of
$\alpha_{\vec{k}}$ and
$\beta_{\vec{k}}$
are found by matching the quantum field and its time derivative at
the transition.
Through this matching, the large time derivative of the massive
mode is converted into a large amplitude of the massless mode.
A non-zero value of
$\beta_{\vec{k}}$ indicates particle creation. In our case,
$\beta_{\vec{k}} \sim (H/k)^{\nu + 1/2}$ for $k\ll H_{\rm I}$, showing that
long-wavelength modes are created with large amplitude. This occurs
also if $m$ goes to zero smoothly \cite{ddpt}, provided that it does so
non-adiabatically, i.e.\
$|\omega''/2\omega^3 - 3{\omega'}{}^2/4\omega^4|>1$, where
$\omega^2(\tau) = k^2 + (m/H_{\rm I}\tau)^2$.

Let us now consider several models which exhibit this behaviour \cite{future}.
First,
consider inflation with two scalar fields, $s$ real and $\phi$ complex,
where $\phi$ couples minimally to the photon field $A_\mu$, giving it a
mass $m_A^2=2 e^2|\phi|^2$.
The potential
is $V_{s}(s^2) + m_\phi^2 |\phi|^2/2 +\lambda_\phi |\phi|^4/4 -
g s^2 |\phi|^2/2$ where $V_s$ is an increasing function. During inflation,
$s$ and $|\phi|$ decrease as they roll along the curve $|\phi|=
\sqrt{(gs^2-m_\phi^2)/\lambda_\phi}$ until
$s<m_\phi/\sqrt{g}$, after which
$|\phi|=0$. In order for our mechanism to be efficient, we require
$m_A^2\approx 10^{-2} H_{\rm I}^2$ and thus
$g/\lambda_\phi \approx 10^{-12}$. Such values arise naturally in supergravity
models,\footnote{We thank A.\ Riotto and W.\ Porod for
independently pointing this out.}
where $g\sim \alpha H_{\rm I}^2/M_{\rm P}^2$, $\alpha\approx 0.1$.

The second possibility is through the back reaction of the vacuum
fluctuations of a scalar field, $\langle \Phi^\dagger\Phi\rangle$,
on the equation of motion of a minimally coupled gauge field. We know that
$\langle \Phi^\dagger\Phi\rangle$ grows during inflation, and thus would
give a mass $2 e^2 \langle\Phi^\dagger\Phi\rangle$ to the photon. The mass
goes to zero if $\Phi$ decays soon after the end of inflation,
which would be the case if $\Phi$ is, e.g., a heavy squark field, or
if $\Phi$ is the electroweak Higgs field and the electroweak
symmetry is restored by reheating. In the latter case,
the electroweak $Z$ field would play the role of the gauge field.
There is some controversy about this; in a different talk at this conference,
Shaposhnikov argues that the effective transverse (so-called magnetic)
mass of the gauge field is zero both during inflation and after inflation.
However, when a quantum operator
evolves non-adiabatically, one should employ out-of-equilibrium methods to
calculate its expectation value, and there is no reason that it should
be zero.

All these mechanisms result in a gauge-field
spectrum ${\cal A}_{\vec{k}}\propto k^{-1-\nu}\sim k^{-3/2}$,
corresponding to a magnetic field
$B_\ell\propto \ell^{-3/2+\nu}
\sim \ell^{-1}$, where $\ell$ is the relevant coherence scale and $\nu\approx
1/2$.
This spectrum is displayed below. Pairs of parallel lines indicate spectra
with and without amplification by a factor $10^5$ during preheating.
We see that the lower bounds for the dynamo seed field can be met with our
mechanism.

\begin{figure}[h!]
\vspace*{-3mm}
\centerline{\epsfxsize = 4in  \epsffile{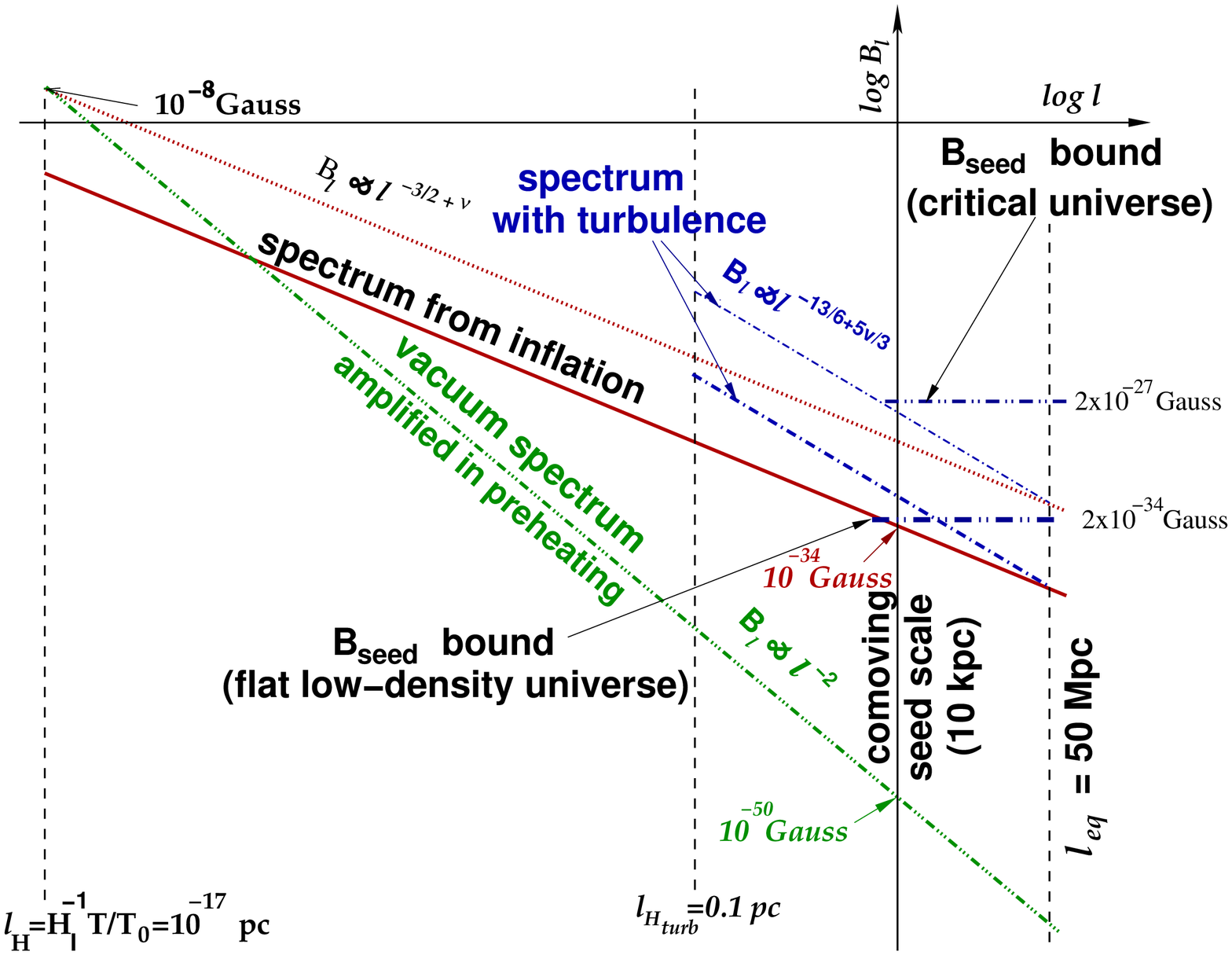}}
\vspace*{-7mm}
\label{tri_fig}
\end{figure}

\end{document}